\documentclass[aps,preprint]{revtex4}%
\usepackage{amsfonts}
\usepackage{amsmath}
\usepackage{amssymb}
\usepackage{graphicx}%
\providecommand{\U}[1]{\protect\rule{.1in}{.1in}}

\begin{document}
\title{Magnetohydrodynamic properties of incompressible Meissner fluids}
\author{A. Maeyens$^{1}$, J.\ Tempere$^{1,2}$}
\affiliation{$^{1}$TFVS, Universiteit Antwerpen, Groenenborgerlaan 171, B2020 Antwerpen, Belgium.}
\affiliation{$^{2}$Lyman Laboratory of Physics, Harvard University, Cambridge MA02138, USA.}
\date{June 18, 2007}

\begin{abstract}
We consider a superconducting material that exists in the liquid state, more
precisely, in which the Meissner-Ochsenfeld effect persists in the liquid
state. First, we investigate how the shape of such a hypothetical Meissner
liquid will adapt to accomodate for an applied external field. In particular,
we analyse the case of a droplet of Meissner fluid, and compute the elongation
of the droplet and its quadrupole frequency as a function of the applied
field. Next, the influence of an applied field on the flow of the liquid is
studied for the case of a surface wave. We derive the dispersion relation for
surface waves on an incompressible Meissner fluid. We discuss some candidate
realizations of the Meissner fluids and for the case of a superconducting
colloid discuss which regime of wave lengths would be most affected by the
Meissner effect.

\end{abstract}
\maketitle

\section{Introduction}

The superconducting materials readily available to man are all solids. Both
for conventional superconductors such as mercury and for high-temperature
superconductors, the critical temperature lies well below the melting
temperature. Nevertheless a crystalline substrate is no necessary prerequisite
for superconductivity. Even for the case of phonon-mediated superconductivity,
it can be argued that liquid metals also exhibit phonons \cite{JaffePRB23}.
Indeed, for the case of dense liquid metallic hydrogen, a superconducting
phase has been predicted \cite{JaffePRB23,BabaevNAT431}.

In this contribution, we investigate how a liquid superconductor would respond
to applied magnetic fields. More precisely, we consider a Meissner liquid,
since we are interested in the magnetic response rather than the electric
response. A candidate realization of a Meissner fluid might be a suspension of
superconducting particles. In Ref. \cite{TaoPRL83,TaoPRB68}, such a suspension
of micron-sized superconducting cuprate particles in liquid nitrogen was
investigated, and when those particles were coated with a layer of ice, acting
as a surfactant preventing the particles from coagulating, a Meissner liquid
state was reported.

The basic property of the Meissner fluid is its capability to expell magnetic
flux $\mathbf{B}$ from its bulk by supporting a persistent surface electric
current $\mathbf{J}_{s}$. In type II\ superconductors above the first critical
field, the flux is incompletely expelled. Here, we consider type I
superconductors or type II superconductors below the first critical field. In
Ref. \cite{LiuJLTP126}, it was shown that when equilibrium is disrupted,
$\mathbf{B}$ and $\mathbf{J}_{s}$ relax back to (new) equilibrium values with
a characteristic time of $\tau=10^{-15}$ s. The time scale for the
hydrodynamic motion of a fluid surface (as determined by the fluid's density
and surface tension) is much slower, and this allows to separate the dynamics
of the currents and magnetic fields from that of the fluid surface. Hence, for
the different problems investigated in the present study, we assume that at
each time the magnetic field has reached its equilibrium value for the given
surface deformation of the fluid \cite{LiuPRL70}.

We start in Sec. II with the study of the shape deformation of a droplet of
incompressible Meissner liquid placed in a uniform external magnetic field.
From the energetics of the optimal shape we calculate the quadrupole mode
oscillation frequency. In Sec. III, we investigate the higher-mode frequencies
by neglecting curvature and modeling surface waves on a Meissner liquid. We
derive the modification of the wave dispersion by a magnetic field parallel
with the surface.

\section{Meissner droplet}

In the absence of gravity, a fixed volume of liquid will form a spherical
droplet to minimize surface tension. When a magnetic field is applied on a
droplet of Meissner fluid, the fluid will magnetize in order to expell the
magnetic flux.\ The energy associated with this magnetization is expected to
be smallest for a cigar-shaped droplet with the axis of cylindrical symmetry
along the magnetic field. However, deforming the droplet into a prolate
spheroid will also increase the surface tension energy. The optimal shape of
the droplet can be found by balancing the magnetic and surface tension energies.

As possible shapes we will consider prolate spheroids with the major axis
parallel to the applied magnetic field. The variational parameters determining
the droplet shape are $a$, the equatorial radius or semiminor axis, and $c$,
the semimajor axis. We denote the applied magnetic field by $\mathbf{H}_{a}$.
The droplet responds to the external field by magnetizing. The magnetization
field $\mathbf{M}$ is zero outside and nonzero inside the droplet. Since then
$\mathbf{\nabla\cdot M}$ is nonzero at the droplet surface, magnetic charges
are induced on the surface. These charges give rise to a demagnetization field
$\mathbf{H}_{d}$.

Inside the droplet, the demagnetization field is straightforwardly related to
the magnetization through $\mathbf{H}_{d}^{in}=-n(a,c)\mathbf{M}$ where
$n(a,c)$ is the demagnetizing factor for a prolate spheroid\cite{BirchEJP6}:
\begin{equation}
n(a,c)=\left(  1-\frac{c^{2}}{\alpha^{2}}\right)  \left[  1-\frac{c}{2\alpha
}\ln\left(  \frac{c+\alpha}{c-\alpha}\right)  \right]  ,
\end{equation}
where $\alpha=\sqrt{c^{2}-a^{2}}$ is the focus distance. The total magnetic
flux is
\begin{equation}
\mathbf{B}=\mu(\mathbf{H}_{a}+\mathbf{H}_{d}+\mathbf{M}),
\end{equation}
with $\mu$ the vacuum permeability. Since inside the Meissner fluid \textbf{B}
has to be zero, we find that the required magnetization satisfies%
\begin{equation}
\mathbf{M}=-\frac{1}{1-n(a,c)}\mathbf{H}_{a}, \label{magnetiz}%
\end{equation}
in agreement with the result of Ref. \cite{CapePR153}.

\begin{figure}
[tbh]
\begin{center}
\includegraphics[
height=2.6364in,
width=6.75in
]%
{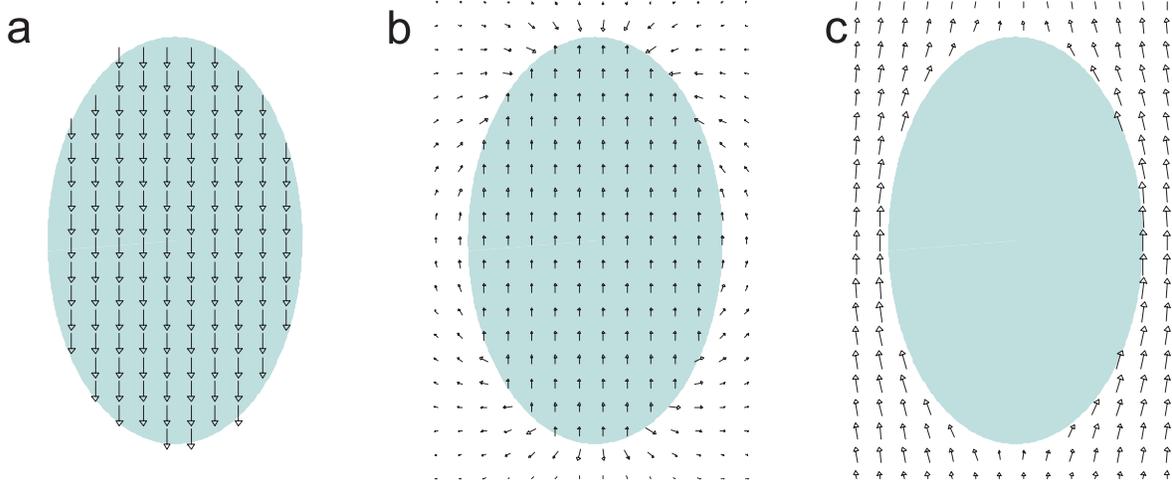}%
\caption{(color online, 2 column) Panel (a) shows the magnetization, which is
only present inside the spheroidal droplet (shaded area), and which is equal
to $-n\mathbf{H}_{a}$. The demagnetisation field $\mathbf{H}_{d}$ is shown in
panel (b). The total magnetic flux $\mathbf{B}=\mu(\mathbf{H}_{a}%
+\mathbf{H}_{d}+\mathbf{M})$ is shown in panel (c). All arrows in panels (a)
and (c) are scaled by the same factor, those in panel (b) are scaled by twice
that amount for visibility.}%
\label{fig1}%
\end{center}
\end{figure}

To find the demagnetization field $\mathbf{H}_{d}^{out}$ outside the droplet,
we solve $\mathbf{\nabla}\times\mathbf{H}_{d}^{out}=0$ (since there are no
induced currents in the volume outside the droplet). This implies that the
outside demagnetization field can be written as the gradient of a scalar
magnetic potential $\mathbf{H}_{d}^{out}\mathbf{=-\nabla}\psi.$ The boundary
conditions are that $\mathbf{H}_{d}\rightarrow0$ at infinity and that the
tangential component of $\mathbf{H}_{d}$ is continuous along the boundary of
the droplet. Equivalently, we can use that the normal component of the
magnetic flux is continuous accross the boundary. The natural coordinate
system to express these boudary conditions are the prolate spheroidal
coordinates \{$\eta,\theta,\phi$\}. The droplet boundary is then defined by
fixing $\eta=\eta_{b}$. The equation for the scalar magnetic potential becomes
\cite{Moon}%
\begin{align}
&  \frac{1}{\alpha^{2}(\sinh^{2}\eta+\sin^{2}\theta)}\left[  \frac
{\partial^{2}\psi}{\partial\eta^{2}}+\coth\eta\frac{\partial\psi}{\partial
\eta}+\right. \nonumber\\
&  \left.  +\frac{\partial^{2}\psi}{\partial\theta^{2}}+\cot\theta
\frac{\partial\psi}{\partial\theta}\right]  =0.
\end{align}
The solution of this equation can be found by separation of variables,
yielding
\begin{equation}
\psi\left(  \eta,\theta\right)  =\overset{\infty}{\underset{n=0}{\sum}}%
A_{n}Q_{n}(\cosh\eta)P_{n}(\cos\theta),
\end{equation}
where $P_{n}$ and $Q_{n}$ are the Legendre functions of the first and second
kind, and the $A_{n}$ are integration constants. The boundary condition
$\mathbf{H}_{d}\rightarrow0$ at infinity has already been used. The second
boundary condition is
\begin{equation}
\left.  \mathbf{H}_{d}^{out}\cdot\mathbf{e}_{\eta}\right\vert _{\eta=\eta_{b}%
}=\left.  \left[  1-n(a,c)\right]  \mathbf{M}\cdot\mathbf{e}_{\eta}\right\vert
_{\eta=\eta_{b}},
\end{equation}
where $\mathbf{e}_{\eta}$ is the unit vector in the $\eta$ direction. This
corresponds to%

\begin{align}
&  -\left.  \mathbf{e}_{\eta}\mathbf{\cdot\nabla}\overset{\infty}%
{\underset{n=0}{\sum}}A_{n}Q_{n}(\cosh\eta)P_{n}(\cos\theta)\right\vert
_{\eta=\eta_{b}}\nonumber\\
&  =H_{a}\frac{\sinh\eta_{b}\cos\theta}{\sqrt{\sinh^{2}\eta_{b}+\sin^{2}%
\theta}}.
\end{align}
We find $A_{n\neq1}=0$ and%
\begin{equation}
A_{1}=-H_{a}\alpha\left[  \frac{c\alpha}{a^{2}}-\ln\left(  \sqrt
{\frac{c+\alpha}{c-\alpha}}\right)  \right]  ^{-1}.
\end{equation}
The demagnetisation field outside the droplet is then%
\begin{align}
\mathbf{H}_{d}^{out}\mathbf{(}\eta,\theta\mathbf{)}  &  =\frac{-H_{a}}{\left[
\frac{c\alpha}{a^{2}}-\ln\left(  \sqrt{\frac{c+\alpha}{c-\alpha}}\right)
\right]  \sqrt{\sinh^{2}\eta+\sin^{2}\theta}}\nonumber\\
&  \times\left\{  \cos\theta\left[  -\coth\eta+\ln\left(  \coth\frac{\eta}%
{2}\right)  \sinh\eta\right]  \mathbf{e}_{\eta}\right. \nonumber\\
&  \left.  +\sin\theta\left[  1+\cosh\eta\log(\tanh\frac{\eta}{2})\right]
\mathbf{e}_{\theta}\right\}  . \label{demagn}%
\end{align}
Fig. \ref{fig1} depicts the various contributions to the magnetic flux: the
magnetization, the demagnetization field, and the magnetic flux. The magnetic
flux remains tangential to the surface, flowing around the boundary of the droplet.%

To calculate the energy difference $E_{M}$ between the unmagnetized and the
magnetized droplet of a given, fixed shape with shape parameters $a$ and $c$,
we use the thermodynamic potential $\tilde{U}=U+\mathbf{H\cdot B}$ with $U$
the internal energy, such that $d\tilde{U}=\mathbf{B}\cdot d\mathbf{H}$ and%
\begin{equation}
E_{M}(a,c)=\int\left[  \int_{\mathbf{0}}^{\mathbf{H}_{a}}\mathbf{B(r)}\cdot
d\mathbf{H(r)}\right]  d\mathbf{r.}%
\end{equation}
Using the results (\ref{magn}),(\ref{demagn}) for the fields, the expression
simplifies to
\[
E_{M}(a,c)=\frac{1}{1-n(a,c)}\frac{\mu H_{a}^{2}}{2}V,
\]
where $V=4\pi a^{2}c/3$ is the volume of the droplet. This term would favour,
for a given volume, a more elongated spheroid. It is counteracted by the
surface tension energy $E_{S}=\sigma S$ with $\sigma$ the surface tension and
$S$ the surface of the spheroidal droplet. This evaluates to
\begin{equation}
E_{S}(a,c)=2\pi\sigma\left[  a^{2}+ac\frac{\sin^{-1}\left(  \sqrt
{1-a^{2}/c^{2}}\right)  }{\sqrt{1-a^{2}/c^{2}}}\right]  .
\end{equation}
To find the optimal surface, the total energy $E_{S}+E_{M}$ need to be
minimized as a function of $\{a,c\}$ with the constraint of constant volume.
If $r_{0}$ is the radius of a spherical droplet with volume $V$, we can
introduce dimensionless parameters $\tilde{a}=a/r_{0}$ and $\tilde{c}%
=c/r_{0}.$ The constraint of constant volume then allows to eliminate one of
the variational parameters since $V=4\pi a^{2}c/3=4\pi r_{0}^{3}/3$ leads to
$\tilde{c}=\tilde{a}^{-2}$. Writing the total energy as $\tilde{E}%
=(E_{S}+E_{M})/(2\pi\sigma r_{0}^{2})$ then leads to%

\begin{equation}
\tilde{E}(\tilde{a})=\frac{\Gamma}{1-n(\tilde{a})}+\tilde{a}^{2}+\frac
{\sin^{-1}(\sqrt{1-\tilde{a}^{6}})}{\tilde{a}\sqrt{1-\tilde{a}^{6}}},
\end{equation}
where $\Gamma$ is a dimensionless parameter expressing the applied magnetic
field
\begin{equation}
\Gamma=\frac{\mu H_{a}^{2}r_{0}}{3\sigma}%
\end{equation}
and
\begin{equation}
n(\tilde{a})=-\frac{\tilde{a}^{6}}{1-\tilde{a}^{6}}\left[  1-\frac{1}%
{2\sqrt{1-\tilde{a}^{6}}}\ln\left(  \frac{1+\sqrt{1-\tilde{a}^{6}}}%
{1-\sqrt{1-\tilde{a}^{6}}}\right)  \right]  .
\end{equation}
Note that $\tilde{a}=1$ corresponds to the spherical droplet, and we need to
find $0<\tilde{a}<1$.

\begin{figure}
[ptb]
\begin{center}
\includegraphics[
height=2.6489in,
width=3.3754in
]%
{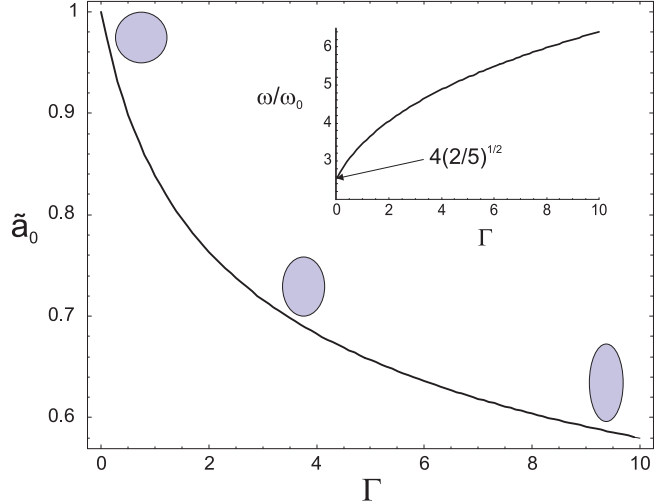}%
\caption{(color online) The droplet shape, characterized by the semiminor axis
expressed in units $r_{0}$ of the spherical droplet with the same volume, is
shown as a function of the magnetic field, expressed through $\Gamma=\mu
H_{a}^{2}r_{0}/(2\sigma)$. In the inset, the oscillation frequency of the
quadrupole mode (in units of $\omega_{0}=\sqrt{3\sigma/(2\rho r_{0})}$) is
shown as a function of $\Gamma$. The limiting value for a spherical droplet is
given.}%
\label{fig2}%
\end{center}
\end{figure}

In Fig. \ref{fig2} the value of $\tilde{a}$ that minimizes the energy,
$\tilde{a}_{0}$, is plotted as a function of the magnetic field in reduced
units. The surface tension of a Meissner liquid contains a contribution
$\sigma_{SC}$ coming from the interface between the superconductor and the
normal state; this should be proportional to the difference between the
coherence length and the penetration depth. There is also a contribution
$\sigma_{L}$ from the liquid-vapour interface. For the Meissner liquid
reported in Ref. \cite{TaoPRL83}, the surface tension is estimated to be of
the order of $\sigma=10^{-3}$ J/m$^{2}$. For a 1 mm droplet, the applied
magnetic flux corresponding to $\Gamma=1$ is then of the order millitesla.
Thus, fields reasonably below the critical field will induce a non-negligible
deformation of the droplet. The smaller the droplet, the more it resists shape deformations.%

A Taylor expansion of the energy around the minimum allows to find the
oscillation frequency of the quadrupole mode of the droplet. We equate the
second order term in the expansion with a harmonic oscillation around the
optimal shape,
\begin{align}
\rho V\omega^{2}  &  =2\pi\sigma\left.  \frac{\partial^{2}\tilde{E}}%
{\partial\tilde{a}^{2}}\right\vert _{\tilde{a}_{0}}\nonumber\\
&  \Rightarrow\omega=\sqrt{\frac{3\sigma}{2\rho r_{0}^{3}}\left.
\frac{\partial^{2}\tilde{E}}{\partial\tilde{a}^{2}}\right\vert _{\tilde{a}%
_{0}}}.
\end{align}
Here $\rho$ is the density of the Meissner liquid$.$ The dependence of the
frequency on the magnetic field in reduced units, $\Gamma$, is shown in the
inset of Fig. \ref{fig2}. The frequency is expressed in units $\omega
_{0}=\sqrt{3\sigma/(2\rho r_{0})}.$ In the limit of spherical droplets
($H_{a}\rightarrow0$) the frequency converges to $\omega/\omega_{0}%
=4\sqrt{2/5}$. For a $r_{0}=$1 mm droplet with $\sigma=10^{-3}$ J/m$^{2}$ and
$\rho=10^{3}$ kg/m$^{3}$, the unit $\omega_{0}$ corresponds to a frequency of
39 Hz. Increasing the magnetic field deforms the bubble, and also stiffens the
oscillation frequency.

\section{Surface waves on a Meissner fluid}

The quadrupole oscillation mode of the droplet is one particular realization
of surface waves. In this section, we investigate how the surface waves on an
infinitely deep Meissner fluid are influenced by an applied magnetic field
parallel to the surface and parallel to the propagation direction of the wave
(a field normal to the propagation direction of the wave would be parallel to
the wave fronts and would not induce magnetic charges on the surface - thus it
would not alter the wave dispersion).
\begin{figure}
[th]
\begin{center}
\includegraphics[
height=4.2869in,
width=3.3754in
]%
{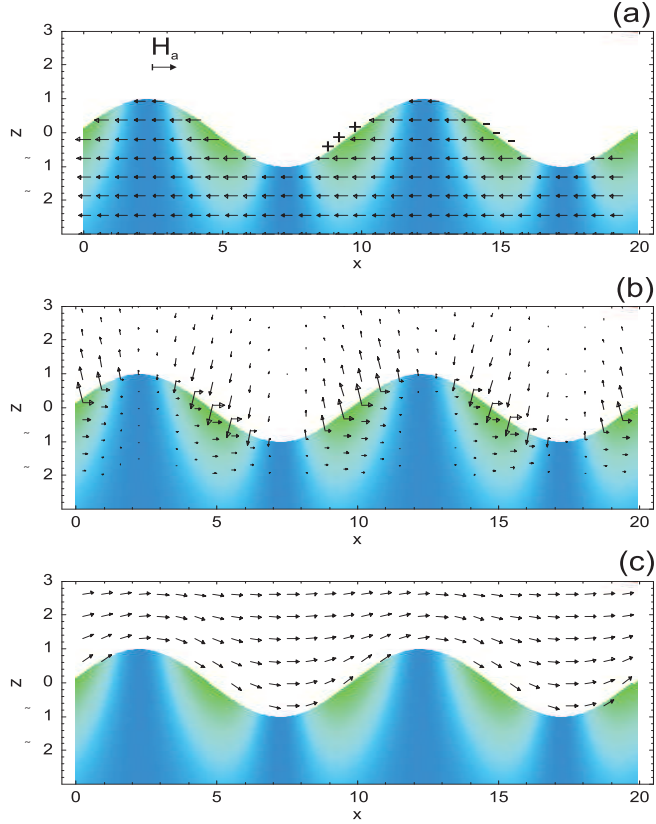}%
\caption{(color online) The magnetization of a Meissner fluid with a surface
wave is shown in panel (a); the applied field is parallel to the surface and
to the wave propagation direction. The magnetization induces surface charges
on the rising and descending wave slopes, as indicated. The shading (from dark
blue - low to light green - high) indicates the concentration of magnetic
charge. In panel (b) and (c), the demagnetization field and the total magnetic
flux are shown, respectively.}%
\label{fig3}%
\end{center}
\end{figure}
Here, we consider small-amplitude waves. The surface of the liquid is
characterized by $\zeta(x)=\alpha\sin(kx)$ where the flat surface corresponds
to the $xy$-plane (at $z=0$), $\alpha$ is the wave amplitude, and
$k=2\pi/\lambda$ is the wave number. The angle between the tangent to the
surface and the horizontal plane is
\begin{equation}
\theta(x)=\arctan[\zeta^{\prime}(x)]=\alpha k\cos(kx)+\mathcal{O}(\alpha^{3}).
\end{equation}

\subsection{Demagnetization field}

When the surface of the fluid is flat and the applied field $\mathbf{H}_{a}$
is parallel to the surface, the magnetization exactly cancels $\mathbf{H}%
_{a}=H_{a}\mathbf{e}_{x}$ to achieve the Meissner state. However, when a wave
is present, the magnetization will induce surface magnetic charge on the
rising and descending slopes of the wave, as illustrated in panel (a) of Fig.
(\ref{fig3}). The shading of the liquid in this figure also illustrates the
concentration of this magnetic charge $\mathbf{\nabla\cdot M},$ strongest at
the slopes of the wave. As in a regular magnetized object, this gives rize to
a demagnetization field $\mathbf{H}_{d}$, calculated below and shown in panel
(b) of Fig. (\ref{fig3}). The resulting total flux is shown in panel (c) of
that Fig. (\ref{fig3}), and, as calculated below, follows the contour of the
surface wave.

To find the magnetic field $\mathbf{H}(x,z)$ in this case, we decompose this
field in a component $H_{//}$ tangential to the surface and a component
$H_{\bot}$ normal to the surface. For a small-amplitude wave, $H_{\bot}$ will
be of order $\alpha$ and $H_{//}\approx|H_{a}|$ up to second order in $\alpha
$. The demagnetization field $\mathbf{H}_{d}=\mathbf{H}-H_{a}\mathbf{e}_{x}$
at the surface of the wave can then be written as%
\begin{align}
&  \mathbf{H}_{d}\left[  x,\zeta(x)\right] \nonumber\\
&  =\left\{  H_{//}\cos\left[  \theta(x)\right]  -H_{\bot}(x)\sin\left[
\theta(x)\right]  -H_{a}\right\}  \mathbf{e}_{x}\nonumber\\
&  +\left\{  H_{//}\sin[\theta(x)]+H_{\bot}(x)\cos[\theta(x)]\right\}
\mathbf{e}_{z},
\end{align}
or, expanding with respect to $\alpha$,
\begin{align}
&  \mathbf{H}_{d}\left[  x,\zeta(x)\right] \nonumber\\
&  =\left[  -H_{a}\frac{1}{2}\alpha^{2}k^{2}\cos^{2}(kx)-H_{\bot}(x)\alpha
k\cos(kx)\right]  \mathbf{e}_{x}\nonumber\\
&  +\left[  H_{a}\alpha k\cos(kx)+H_{\bot}(x)\right]  \mathbf{e}%
_{z}+\mathcal{O}(\alpha^{3}).
\end{align}
For the calculation of the energy associated with the magnetization, we need
to find the demagnetization field inside the fluid. As a trial solution, we
use%
\begin{equation}
\mathbf{H}_{d}^{in}(x,z)=\mathbf{h}(x)e^{k(z-\zeta(x))}. \label{vormpje}%
\end{equation}
That is, we assume an exponential decay into the fluid with the wave length of
the surface wave as a characteristic scale. In the direction of propagation of
the wave, we need to find $\mathbf{h}(x)=h_{x}(x)\mathbf{e}_{x}+h_{z}%
(x)\mathbf{e}_{z}.$ The ansatz (\ref{vormpje}) allows to satisfy $\nabla
\times\mathbf{H}_{d}^{in}=0,$ which reduces to
\begin{equation}
\frac{\partial h_{z}(x)}{\partial x}-h_{z}(x)\alpha k^{2}\cos(kx)=kh_{x}(x),
\end{equation}
with boundary conditions $\mathbf{h}(x)=\mathbf{H}_{d}\left[  x,\zeta
(x)\right]  $ and $\mathbf{h}(x)=\mathbf{h}(x+\lambda)$. This can be solved
straightforwardly, yielding%
\begin{align}
\mathbf{H}_{d}\left[  x,z<\zeta(x)\right]   &  =H_{a}\frac{\alpha^{2}k^{2}}%
{2}\cos^{2}(kx)\nonumber\\
&  \times\exp\{k\left[  z-\alpha\sin(kx)\right]  \}\mathbf{e}_{x}.
\end{align}
for the demagnetization field inside the liquid.

\subsection{Magnetic energy}

Consider a surface of fluid of length $L_{y}$ in the $y$-direction and
$L_{x}=p\lambda$ ($p\in\mathbb{N})$ in the $x$-direction. The energy
difference $E_{mag}$ between the unmagnetized ($\mathbf{H}_{a}=0$) and the
magnetized case can be written as%
\begin{equation}
E_{mag}=\frac{\mu}{2}\underset{z<\zeta(x)}{\int}\mathbf{H}_{a}\cdot
\mathbf{H}_{d}d^{3}\mathbf{r}+\frac{\mu}{2}\int\mathbf{H}_{a}^{2}%
d^{3}\mathbf{r}.
\end{equation}
The second term represents the energy of swichting on the applied magnetic
field, and does not depend on the presence of a surface wave. The first term
evaluates to
\begin{align*}
E_{mag}  &  =\frac{\mu H_{a}^{2}}{2}\frac{\alpha^{2}k}{2}L_{y}\int_{0}^{L_{x}%
}dx\cos^{2}(kx)\\
&  =\frac{1}{4}\frac{\mu H_{a}^{2}}{2}k\text{ }L_{x}L_{y}\ \alpha^{2}.
\end{align*}
Thus, the magnetic energy per unit surface required to establish the Meissner
state when there is a surface wave with wave number $k$ and amplitude $\alpha$
is (to order $\alpha^{3}$) given by%
\begin{equation}
E_{M}=\frac{\mu H_{a}^{2}}{8}k\alpha^{2}. \label{EM}%
\end{equation}

\subsection{Wave dispersion}

To find the dispersion relation of a surface wave on an infinitely deep
Meissner fluid, we follow the Hamiltonian procedure outlined in Ref.
\cite{KlemensAJP52}. The kinetic energy per unit surface, associated with the
surface wave is
\begin{equation}
E_{K}=\frac{1}{4}\frac{\rho}{k}\dot{\alpha}^{2}. \label{EK}%
\end{equation}
When the flat surface is deformed, restoring forces tend to pull it flat
again. These restoring forces can be related to the surface tension energy,
per unit surface:%
\begin{align}
E_{S}  &  =\frac{1}{L_{x}}\overset{L_{x}}{\underset{0}{\int}}\frac{1}{2}%
\sigma\left(  \frac{\partial\zeta(x)}{\partial x}\right)  ^{2}dx\nonumber\\
&  =\frac{1}{4}\sigma k^{2}\alpha^{2}, \label{ES}%
\end{align}
and to the gravitational energy, per unit surface:
\begin{equation}
E_{G}=\frac{1}{4}\rho g\alpha^{2}. \label{EG}%
\end{equation}
In a Meissner liquid, subjected to a magnetic field parallel to the surface
wave propagation direction, also the magnetic energy (\ref{EM}) gives rise to
a restoring force. The (classical) Hamiltonian associated with the surface
wave of amplitude $\alpha$ is then given by the sum of\ (\ref{EK}%
),(\ref{ES}),(\ref{EG}) and (\ref{EM})\ :%
\begin{equation}
H=\frac{1}{4}\frac{\rho}{k}\dot{\alpha}^{2}+\left(  \frac{1}{4}\sigma
k^{2}+\frac{1}{4}\rho g+\frac{1}{4}\frac{\mu H_{a}^{2}}{2}k\right)  \alpha
^{2}.
\end{equation}
This expression is valid for small-amplitude oscillations, and leads to a
dispersion relation%
\begin{equation}
\omega(k)=\sqrt{\frac{\sigma k^{3}}{\rho}+\frac{\mu H_{a}^{2}}{2}\frac{k^{2}%
}{\rho}+gk}.
\end{equation}
The effect of the magnetic field dominates when $\mu H_{a}>>\sqrt{\mu\rho
g/k}$ and $\mu H_{a}>>\sqrt{\mu\sigma k}.$ Taking again $\sigma=10^{-3}$
J/m$^{2}$ and $\rho=10^{3}$ kg/m$^{3}$, and a magnetic field of $\mu
H_{a}=100$ G, we find that the Meissner contribution to the dispersion
dominates for $10^{2}$ m$^{-1}\lesssim k\lesssim10^{5}$ m$^{-1}.$ The
experimental setup of Ref. \cite{TaoPRB68} would allow to probe this range of
wave lengths. Surface waves can be detected by reflecting a laser beam off the
surface of the Meissner fluid and noting the displacement of the laser spot as
a function of time and space.

\section{Conclusion}

Fluids with remarkable magnetic response properties, such as ferrofluids, have
sparked a lot of interest. Here, we investigate how a fluid superconductor, a
Meissner fluid, would react to an applied magnetic field. Even though such
Meissner fluids are not yet accessible, candidate realizations from theory
\cite{BabaevNAT431} and recent experiments \cite{TaoPRL83} can be found.

When such a fluid superconducting material is placed in an applied magnetic
field, it will flow to adapt its shape and minimize the energy required to
expell the magnetic flux. In this contribution, we focused on a single droplet
of Meissner fluid and derived the deformation of such a droplet due to an
applied magnetic field as well as the associated quadrupole oscillation
frequency. Inversely, imposing a hydrodynamic flow, for example by generating
a surface wave, will alter the energetics and thus the dispersion of the wave.
We calculated how the dispersion of the surface wave is changed due to the
presence of a magnetic field, applied parallel to the surface and to the
propagation direction of the wave. The change in the dispersion is predicted
to be relevant for the wave length range thought to be achievable in a setup
such as \cite{TaoPRL83}.

\begin{acknowledgments}
Acknowledgements -- Discussions with M.\ Wouters and J.T. Devreese are
gratefully acknowledged. This work has been supported in part by FWO-V
projects Nos. G.0356.06, G.0115.06 and G.0435.03. J.T. gratefully acknowledges
support of the Special Research Fund of the University of Antwerp, BOF\ NOI UA 2004.
\end{acknowledgments}


\bigskip

\begin{thebibliography}{99}                                                                                               %


\bibitem {JaffePRB23}J. E. Jaffe and N. W. Ashcroft, Phys. Rev. B \textbf{23},
6176 (1981).

\bibitem {BabaevNAT431}E. Babaev, A. Sudb\o , and N. W. Ashcroft, Nature
\textbf{431}, 666 (2004).

\bibitem {TaoPRL83}R. Tao, X. Zhang, X. Tang, and P. W. Anderson, Phys. Rev.
Lett. \textbf{83}, 5575 (1999).

\bibitem {TaoPRB68}R. Tao, X. Xu, and E. Amr, Phys. Rev. B \textbf{68}, 144505 (2003).

\bibitem {LiuJLTP126}M. Liu, Journ. Low.\ Temp. Phys. \textbf{126}, 911 (2002).

\bibitem {LiuPRL70}M. Liu, Phys. Rev. Lett. \textbf{70}, 3580 - 3583 (1993).

\bibitem {BirchEJP6}C. Birch, Eur. J. Phys. \textbf{6}, 180\textbf{ }(1985).

\bibitem {CapePR153}J. A. Cape, J. M. Zimmerman, Phys. Rev. \textbf{153}, 416 (1967).

\bibitem {Moon}P. Moon and D.E. Spencer, \emph{Field Theory Handbook}
(Springer-Verlag, 1971), p. 28-36.

\bibitem {KlemensAJP52}P.G. Klemens, Am. J. Phys. \textbf{52}, 451 (1984).
\end{thebibliography}
\end{document}